# Governance and Communication of Algorithmic Decision Making:

## A Case Study on Public Sector

**Eric Jonk**, Department of Information Sciences, Open Universiteit, Heerlen, the Netherlands

**Deniz Iren**, Center for Actionable Research of the Open Universiteit and Department of Information Sciences, Open Universiteit, Heerlen, the Netherlands

Contact: jonk@ziggo.nl

# Governance and Communication of Algorithmic Decision Making: A Case Study on Public Sector

Eric Jonk
Department of Information Sciences
Open Universiteit
Heerlen, the Netherlands
jonk@ziggo.nl

Deniz Iren
Center for Actionable Research of the Open Universiteit
and Department of Information Sciences
Open Universiteit
Heerlen, the Netherlands
deniz.iren@ou.nl

*Abstract*— **Algorithmic Decision Making (ADM) has permeated all aspects of society. Government organizations are also affected by this trend. However, the use of ADM has been getting negative attention from the public, media, and interest groups. There is little to no actionable guidelines for government organizations to create positive impact through ADM. In this case study, we examined eight municipal organizations in the Netherlands regarding their actual and intended use of ADM. We interviewed key personnel and decision makers. Our results show that municipalities mostly use ADM in an ad hoc manner, and they have not systematically defined or institutionalized a data science process yet. They operate risk averse, and they clearly express the need for cooperation, guidance, and even supervision at the national level. Third parties, mostly commercial, are often involved in the ADM development lifecycle, without systematic governance. Communication on the use of ADM is generally responsive to negative attention from the media and public. There are strong indications for the need of an ADM governance framework. In this paper, we present our findings in detail, along with actionable insights on** *governance*, *communication*, **and** *performance evaluation* **of ADM systems.**

## I. Introduction

Digital transformation in every sector is currently driven by ever-growing adoption of Algorithmic Decision Making (ADM). Regardless of the form it takes and the name it is given (e.g., ADM, data science, machine learning, AI, big data), algorithms that rely on a large amount of data to make predictions, classifications, or inferences are heavily being used in decision making in every aspect of human society. The benefits of ADM are apparent, as they offer strategic opportunities, increased productivity [1], and increased speed and efficiency [2].

Despite the potential benefits [3], [4], the use of ADM is criticised due to a lack of transparency and accountability, the biases that may hinder fairness [5], and the difficulty to explain their working procedure and the outcomes. There is a reluctance to trust ADM; people lose confidence in ADM more than they lose in humans even when they make the same mistakes [6]. In a 2019 study [7], 80% of Dutch citizens stated that they want supervision on the use of algorithms, while only 20% think that ADM leads to a safe and healthy world and 12% think that the use of ADM contributes to the fairness and justice of the system.

Public concerns for ADM cause additional challenges for government organizations to adopt ADM since the core principles of governments are based on fairness, transparency, and accountability. On the one hand, public administration strives to achieve an increased efficiency in their operations, on the other hand, they have the obligation to protect their citizens from the potential risks of ADM [8].

Governments require new capacity and adjustment in laws to cope with these new challenges [9]. Since the technological and societal changes outpace the changes within public administrations, a proactive approach is needed [10].

ADM has a large potential to improve the public services in municipalities. For many public services, the municipality is the first and primary point of contact for citizens. Service delivery may improve for the population in general, but it could be possible that some (groups of) citizens experience undesired adverse effects. Hence, the goal of our research is to shed some light on how local government organizations can ensure that ADM has positive impacts on their citizens, and to provide guidance towards the development of a prescriptive ADM governance framework in the future. In our attempt to explore various aspects regarding the actual and the envisioned use of ADM, we define four research questions as follows.

- How is ADM currently being used by municipalities, and what is the envisioned use of ADM in the future?
- What is the actual and envisioned governance mechanism of ADM in local governments?
- What is the policy for communicating the use of ADM with citizens?
- How is impact/performance of ADM measured?

To address these questions, in this study, we conducted a multiple case study on eight municipal organizations in the Netherlands. We applied archival study and explorative, in-depth, semi-structured interviews with the officials from the selected organizations. Our findings paint a rich picture of the state of municipal organizations on ADM. We formulate our findings as actionable insights to governments, municipal organizations, researchers, and citizens for ADM use.

The remaining of this paper is structured as follows. Chapter 2 covers our findings from the literature review. Chapter 3 describes our research method. Chapter 4 elaborates on the results. In Chapter 5 we discuss the results, provide actionable insights, conclude our paper.

## II. Background and Related Work

### A. Background

**Traditional information systems (IS) and ADM**: Traditional IS embody hardcoded logic and rules that could be followed step by step. The outcomes of such systems can be predicted and traced back to the inputs. This makes controlling and auditing possible, thus, enables accountability. However, the advances in computation technology and the abundance of data paved the way for the evolution of traditional IS into ADM systems. Two key characteristics of ADM are the essential dependence on data and the increased complexity. The behaviour of an ADM model entirely depends on the data that are used to train it. When combined, these two characteristics make it especially challenging to explain (or even understand) their operational details and how they reach their outcomes [11]. Most organizations that use ADM entirely rely on their proper functioning without knowing the "ins and outs".

In this paper we use the terms *ADM* and *data science* to cover all IS and analytics that rely on data to make classifications, predictions, and associations to be used in decision making. These include machine learning, AI, neural networks, certain types of data analyses and business intelligence, data science models, and big data analytics.

**Concerns about ADM use**: Despite the apparent benefits, the use of ADM raises many serious concerns. Several prominent concerns are (un)fairness, biases, and algorithmic discrimination [12]; unclear accountability [13]; a lack of trust in algorithms [6]; and limited transparency and explainability [14].

*Fairness, biases, and algorithmic discrimination*: Discrimination is defined as the application of different rules to similar cases [15]. ADM runs the risk of discrimination because it reflects the biases that already exist in the society. Several sources for biases are identified by researchers [16] such as biased historical data to train algorithms [17], imbalanced representation of cases in the data, misuse of classification models in the wrong contexts, and the use of the same dataset for training the model and testing its performance [5].

*Unclear accountability*: Accountability is among the core values of any democratic government [18], [19], which makes it an essential consideration for the use of ADM in public administration. However, there is significant public concern that ADM is too opaque to provide accountability [13]. Governments are expected to avoid the potentially harmful impacts of ADM on their citizens while harnessing the benefits of algorithms [8]. At the current state, ADM systems are mostly unregulated even as they dictate power over the citizens and policies [20]. Methods have been suggested by researchers to attain ADM accountability [21], [22] such as improving transparency [20] and defining regulation and governance methods [23]. Moreover, rethinking the evaluation of ADM to cover not only accuracy, effectiveness, and efficiency but also testability and auditability [24] could be a step toward algorithmic accountability.

*Lack of trust in algorithms*: The opaque nature and data privacy considerations hinder trust in ADM. Public feel an undesired lack of control over the algorithms that make decisions concerning them [17], which gives another reason to doubt the benevolence of ADM. Moreover, there is an observable algorithm aversion phenomenon: when individuals witness ADM systems making mistakes, they lose much more confidence in them than they would when a human makes the same mistake [6].

*Limited transparency and explainability*: Administrative procedures and decisions by the government should be subject to the opportunity of "objection and appeal", and ultimately the judgment of a court. For this to function, the government should be able to explain what happened in the ADM, and why. GDPR includes a right to explanation clause. However, there are conflicting views on the effectiveness of this clause. Several scholars see GDPR as an incentive to create more fair and transparent algorithms [25] while others emphasize that it does not provide a solid legal mandate to enforce a meaningful right to explanation [26]. The right to explanation clause of GDPR is considered vague and too flexible. Moreover, even with the best intentions, in some cases, it is technically not possible to achieve explainability. These concerns may be alleviated through enhanced transparency and better communication strategies [14].

*B. Related Work*

In a non-academic exploratory study, Doove has conducted a survey on the use of algorithms in the Dutch government organizations [27]. 67 organizations have responded to the survey. 48% of them stated that they use algorithms in their processes. Most of the pilot projects that were undertaken in the early phase of ADM adoption had the goal of identifying high risk cases such as fraud. The participants reported that their choice of algorithms depend on explainability, testability, and accuracy. Respondents had conflicting view regarding the need for algorithm regulation, as half of them desire policies that regulate the use of algorithms, and others want to retain the flexibility at the development phases. This study shows similarities to ours in terms of its objectives. However, our work differentiates itself as it follows a systematic case research method with in-depth semi-structured interviews which enabled us to flexibly explore the use of ADM in local government organizations.

There is recently emerging literature that refers to the use of ADM in public sector. Most studies focus on particular use cases such as, tax fraud detection [28], optimization of waste collection services in a smart city [29], while several studies examine the overall needs and concerns for ADM use in public sector [8], [30], [31].

III. RESEARCH METHOD

We used an exploratory multiple case study design to explore how municipal governments use ADM and how the use of ADM benefits citizens. Our case study design follows the guidelines of Yin [32]. Case research allowed us to gather richer data via semi-structured interviews and archival document analysis. The research follows an inductive approach, i.e., data gathering to explore the phenomenon, and an attempt at structuring this data into a conceptual framework [33].

*A. The sample and the unit of analysis:*

We aimed at selecting municipalities that are representative of all municipalities in terms of certain characteristics. All municipalities are bound by the same law and they provide similar services to citizens. Key difference in their characteristics is the population that they address. Larger municipalities have higher budget. This means they have advantages of scale when investing in innovations for specific systems and competences, such as data analytics. Thus, we decided to include municipalities of various sizes (i.e., small/medium/large). For the purpose of our research, we also aimed at focusing on municipalities that have shown interest in data analytics in the past. Therefore, we shortlisted the municipalities which are in contact with VNG (Vereniging van Nederlandse Gemeenten; Dutch Association of Municipalities) regarding the use of data analytics. The sample consisted of eight municipal organizations that fit the selection criteria. Three cases were small municipalities (population less than 50.000), two were middle-size (population 50.000 – 100.000), two large (population > 100.000) municipalities, and one shared services organization that is providing services for seven municipalities with a total population of almost 300.000.

*B. Data collection:*

First, we studied the following public documents and artefacts: brochures and websites of the selected municipalities and VNG, documents that provide details of governance and work processes, and a generic search on the Internet. Using the initial findings, we determined a set of topics to guide the interviews. Second, we conducted semi-structured interviews using template analysis.

TABLE I. INTERVIEW TEMPLATE

| CODE | Topics |
|---|---|
| ALG1 | The use of ADM |
| ALG1.1 | Policy fields, examples, people involved |
| ALG1.2 | Considerations for using ADM |
| ALG1.3 | Benefits of using ADM |
| ALG1.4 | Risks of using ADMS |
| ALG2 | Developments and plans about the use of ADM |
| GOV1 | Governance of ADM |
| GOV1.1 | Involvement of third parties |
| GOV1.2 | Involvement/awareness of the senior management and the council |
| GOV2 | Developments and plans for governance |
| POL1 | Remarks related to the policy process |
| COM1 | Communication about ADM |
| COM2 | Developments and plans for communication |
| PFM1 | Current performance measurement |
| PFM2 | Developments and plans for performance measurement |
| OTH1 | Other aspects that could determine the impact of ADM on the quality of services |

The interview topics were determined in a way that provides flexibility to explore and were transformed into an interview template (Table 1). The interviews contained a short introduction, an explanation of the research, prior to the questions and answers. The research interview ". . . is about asking purposeful questions and carefully listening to the answers to be able to explore these further" (Saunders et al., 2015). The type of interview should suit the purpose of the research. Since the research was exploratory and inductive, semi-structured in-depth interviews were used. Other reasons for using this type of interview were the range of topics addressed, the open-ended nature of the questions, and the different backgrounds and personalities of the interviewees. The interview protocol was as the following. Interviews start with a short introduction, followed by an explanation of the research, and continues by addressing a list of topics and background information. Each interview was audio-recorded with permission of the interviewee. The interviews were transcribed in full and sent to the interviewee for feedback. All transcriptions were agreed upon by the interviewees. Consecutively, the relevant statements in the transcriptions were coded, using the template. For every coded statement, a short interpretation and analysis was made. The interview format was used as a guideline during the interviews, not necessarily addressing all the questions but adjusting with the flow of the conversation, depending also on the expertise and background of the interviewee. This allowed to explore topics a little further or deeper if the expertise allowed for it, and skipping topics where expertise lacked. For every topic, questions were asked about the current situation, as well as the views and ideas on future developments.

*C. Data analysis*

We applied thematic analysis to analyze the interview transcripts (Saunders et al., 2015). Our approach consisted of the following steps: a) *Familiarizing with the data*: achieved by preparing, conducting, and transcribing the interviews., b) *coding*: labelling units of data, specifically sentences or paragraphs with a code, c) *searching for themes and relationships*: identifying the relevant patterns and their relationships, d) interpreting the themes, and e) refining themes and summarizing the interpretations.

IV. RESULTS

Table 2 shows the description of the eight municipal organizations and the roles of the 12 officials as the interviewees. After coding the interviews, 832 statements were identified as relevant with the research questions of this study. The remaining of this section elaborates on our findings organized in related subheadings.

TABLE II. INTERVIEWEES AND MUNICIPALITIES

| Organization | Size | Role of the interviewee(s) | Notes |
|---|---|---|---|
| M1 | Small | BI specialist | |
| M2 | Large | Data advisor | |
| M3 | Large | Manager of research and data center | Service delivery to municipalities |
| M4 | Large | Senior data science researcher | |
| M5 | Small | Data scientist / BI specialist | |
| M6 | Medium | Data project leader, policymaker, BI specialist | |
| M7 | Medium | Alderman, data scientist, policy advisor | |
| M8 | Small | Data engineer | |

**ALG1: The use of ADMs**

Only one municipality replied that they explicitly use data science. The majority of the rest acknowledges that they undertake data science in the form of business intelligence (BI) that relies on algorithms as a part of 'data driven working' program that is sponsored by VNG. No municipality systematically uses data science yet, meaning; there is no budget, human resources, IT capabilities, etc. explicitly and specifically allocated to conduct data science projects. Four of the municipalities have initiated pilot projects to get acquainted with data science rather than to get results. Data science may already be used within departments of some municipalities, "under the radar". In those cases, third parties are often involved.

### ALG1.1: Policy fields, examples, people involved

When asked about the policy fields in which ADMs are mostly used, the majority of the municipalities name social domain. Within the social domain, welfare and social services are reported as specific policy fields for initiating pilot projects to achieve "data awareness". Particular use case examples are fraud detection and financial forecasting. The ranked non-exhaustive list of the reported policy fields is the following: social domain, environment, economy, tax, and education.

The municipalities primarily rely on internal resources when they venture into ADM initiatives. If internal resources do not exist, they hire experts externally or outsource the project by procurement. The departments that are mostly involved in the ADM initiatives are information management, information technologies, public affairs, and third parties. Rarely, the municipalities mentioned the involvement of policy makers, strategic advisors, and business operatives. Only a few municipalities already employ data scientists and they still state that they lack the expertise to undertake data science projects. Even when data scientists are present, they are mostly working on business intelligence cases and data management. They work closely with business experts, either directly or via business analysts. Occasionally, internal experts such as privacy officers and data protection officers also join the data science initiatives. However, the involvement of senior management is limited to incidental contacts to explore the requirements for possible projects.

### ALG1.2: Considerations for using ADMs

Most municipalities are in the process of improving their BI and data management practices. Introducing data science is seen as the next step. They adopt a healthy innovative mindset by seeing mistakes as a way of learning. However, most of them are troubled by uncertainties about the data, expertise, human resources, the business case, governance, and communication. Furthermore, due to the negative attention to ADMs in the media, they are hesitant to take the next step.

Embedding the use of data science in the organization is a challenge for all. That is also why they are first trying to improve BI. It acquaints the municipalities with new "information products", and creates a basic infrastructure for interaction between the data specialists, management and policy experts of the primary processes. Data science is perceived as the "next level" and currently too complex for this setting, also due to a lack of resources.

"*Without a good business case, not a lot gets off the ground in municipalities*", says one interviewee. All municipalities are intentionally working based on demand and they aim to keep doing so in the near future. The interviewees acknowledge the need to help internal clients to draft appropriate data science questions to define the business case better. However, standalone exploration of data with data science is seen as out of reach.

### ALG1.3: Benefits of using ADMs

Municipalities state that they experienced benefits by improving their BI, using interactive dashboards, better presentation, more and better data, better risk management, and a learning cycle for information management. The process of improving BI is deemed helpful, or even a prerequisite, for initiating data science practices.

In one case, the municipality used ADM to improve the detection and prevention of welfare fraud. They report 95% reduction in time spent by public servants while preserving the effectiveness. Additionally, 95% of the citizens no longer have administrative obligations such as filling out tax declarations. This is considered a significant efficiency benefit to the citizens.

Municipalities undertaking pilots or experiments with data science are struggling to clearly identify the benefits, whilst being convinced of the benefits data science can provide. The problems mentioned are uncertainty about criteria or standards to use, not being able to get a concise question and too much focus on getting results.

### ALG1.4: Risks of using ADM

The risks mentioned fall into four categories: a) General: there is a negative attention in media and politics on privacy, and losing the "human influence" on decision making. b) Organization: knowledge and competencies are scarce, and governance is inadequate. There is also a lack of oversight by management and support from employees. c) Data: data quality is insufficient, and the available data have biases. There is uncertainty about the standards that are in use. Additionally, it is not clear if the risks and benefits are weighed and balanced, who is responsible for doing so. d) Execution: the municipalities have troubles formulating data science questions as there is too much focus on one aspect, for example, getting results. There is little attention for the impact of mistakes.

### ALG2: Developments and plans about the use of ADM

The interviewees expect that ADM will be part of standard applications, and working with data is the future. The use of ADM is believed to increase in society in general. Therefore, municipalities will have to prepare for that. However, the potential for smaller municipalities is not yet clear.

All the municipalities aim at increasing the use of data science. They have plans to adapt their organization, mainly by hiring new staff and introducing governance. Some of their indications are as the following. New rules and regulations should not diminish the added value of algorithms and data science. Creating a safe environment, such as a laboratory, to experiment and to develop tools is important. To increase the use of ADM in a controlled way, data quality should be improved and the employee support should be strengthened. The use of ADM should be done carefully, always keeping humans in the decision-making process. The municipalities should work together towards adopting data science, by sharing their knowledge and experiences. The impact of ADM and potential errors should be carefully considered. It is important to weigh different interests, and to be transparent about the choices and interpretations made in the process.

### GOV1: Governance of ADM

Governance in municipalities currently focuses on financial accountability, and managing legal and technical risks. Most municipalities exercise governance of technical and legal aspects of data and data usage, i.e., storage, authorization, data sharing, meta data, privacy protection, and anonymization. More substantive aspects, such as standards for quality (i.e., accuracy, completeness, and timeliness), actual data content and meaning, and data processing are not yet covered. The focus is on risk reduction, mainly concerning privacy and data leaks.

One municipality that has already been working with an ADM system in the last couple of years states that it is still a process of trial and error, making patchwork solutions as problems arise. Moreover, they are working with a third-party contractor which adds to the complexity. Responsibility for the development and use of data science is generally unclear. Municipalities explicitly state that the information products (e.g., BI systems) are developed by one department and deployed by another. However, neither have formal procedures of governance. The common opinion is that eventually the operational manager is -or should be- responsible for the deployment of ADM and use of the results.

Municipalities acknowledge that using data requires a multi-disciplinary approach. This means that data scientists often work together with the front office employees. Occasionally, policy makers or strategy experts are also engaged. Such multidisciplinary collaboration calls for structural governance. However, the involvement of management is usually bottom up, and not systematic.

On external governance, municipalities are awaiting national initiatives. They express a need for governance or regulation, but are wary of administrative burdens.

### GOV1.1: Involvement of third parties

All municipalities are working with third parties when engaging in data science, believing that it is inevitable. Third parties can be commercial companies that supply services or resources, other public institutions such as Central Bureau of Statistics (CBS), universities, and regional cooperatives. Off-the-shelf software is being used in primary processes of the municipalities. Such software increasingly relies on data.

The reasons for involving third parties are: a lack of own resources and expertise, a better business case, and sharing knowledge. Often the initiative comes from the third party, approaching the municipality with new products or improved services. One municipality mentioned their strategy of 'scanning the market' before attempting specific data science projects on their own. An experienced third party is likely to be cheaper, quicker, and better. Collaborating with a successful third party is seen as a valuable learning opportunity.

Engaging with third parties requires special attention to governance. Although the municipalities have procedures for procurement, these do not cover important aspects of data science, such as risk identification and risk management, accountability, or communication. There are certain cases in which the cooperation does not require any purchasing or financial transactions, thus, in those cases even the procurement procedures are not used. Municipalities are also unsure about the management of third parties when working together on data science. The minimal level of expertise required to oversee the contracted work is often unclear or simply does not exist in the organization. All municipalities agree that cooperation and knowledge sharing should be encouraged and managed by a national institution such as VNG.

### GOV1.2: Involvement/awareness of the senior management and the council

Council and senior management are informed about any use of data science. They are involved in general (strategic) decisions about using data science but this is a "bottom-up" affair. There are hardly any top-down initiatives. Data science is not yet used to improve the strategic decision-making process. Accountability is still traditional, which is mostly about the financials.

Middle management is mainly focused on using data science and BI in their own sector. An overarching vision or strategy from senior management is lacking. In cases when risks materialize which might result in public, political, and media attention on data science, the senior management becomes involved. Such moments trigger ad hoc attention and involvement for important aspects such as governance and communication. Especially, the council and aldermen can improve trust by listening to and communicating with citizens and interest groups.

### GOV2: Developments and plans for governance

The distinction between ADM and traditional software tools needs to be clarified for better governance. There is still unclarity about this distinction in municipalities, resulting in inaction. Even though a few municipalities state that they perceive ADM as merely a tool, others mostly acknowledge that using data science poses new, additional challenges for public organizations, especially for governance.

Municipalities express the need for a governance framework, guidelines, and regulations for ADM. They are expecting these from national level institutions such as the ministries or VNG. This could also include external supervision and inspection. A framework should strike a balance between creating added value and preventing excess, with as little administrative burdens as possible. Despite the unison in acknowledging the need for a governance framework for ADM, the municipalities are in the dark about how to implement and maintain it.

All municipalities find it important to be able to experiment with ADM in order to maximize the added value. This implies accepting mistakes, and learning from them to improve. Municipalities are still unsure about the scale and scope of using ADM, especially what they should and can do themselves, and when to engage with third parties. They want to manage the relationship with the third parties, and to stay in control of the collaborative data science projects but they do not yet embody the sufficient expertise to do so. Therefore, they acknowledge the need for special attention to governance of interaction with third parties, such as procurement or cooperation.

Most municipalities foresee organizational changes to accommodate increased use of data science. This is mostly initiated by the ICT department. The current focus is on the tooling, preconditions, and risks. "*We prefer to thoroughly sort things out, before algorithms are implemented in any process at all*", says one interviewee. Both internal and external communication on the use of ADM in public services are thought to be addressed as part of governance. Gaining trust is an important goal, but a concrete plan for how to achieve this goal still does not exist.

All agree that (senior) management and politicians should be (more) aware of the importance of data for decision making. They should be more aware of the possibilities and risks of using ADM so they can be more involved in the earlier stages of the policy process. On an operational level, the manager of the primary process should be responsible for the deployment of information products; the manager of the business operations department for the development and maintenance. Finally, most municipalities state that when using data science it is important that there is always human involvement: no automated decision making.

**POL1: Remarks related to the policy process**

Due to improved BI (reports, dashboards) the council and management in most municipalities are already better informed. They could be able to make better and quicker decisions. However, since policy processes are still mostly unchanged, the frequency and detail of decision making have not changed yet. Therefore, the adoption of policies to data-centric decision making is a requirement to remove this bottleneck.

Even though laws and regulations are the same for all municipalities, policy processes may differ within the municipality, as well as between municipalities. Every municipality is free in procuring their own ICT systems and support, and in organizing and implementing both primary and secondary processes. This poses a challenge for the development of an ADM governance framework that fits all municipalities.

Finally, several municipalities made interesting remarks about the value of data. Good data (i.e., actual, correct, complete) is seen as an asset; a valuable possession. One municipality is defining ADM models and tools as "information products", in addition to other tools such as BI solutions. These views seem to alleviate the reluctance to use data science in policy processes.

**COM1: Communication about ADM**

Regarding communication about the use of data science, municipalities state vague ambitions to be transparent and open about operating ethical and "doing the right thing". Actual communication (in the form of information presented on the municipal website) is limited to informing citizens about the GDPR, and statements about privacy.

Half have reported the specific mentioning of data science on their website. Only one municipality has any substantial mention of the terms "algorithm" or "data science" on their website. Only one municipality has actual experience with (external) communication about their use of data science. The others state that there may be some internal communication, in the form of training, information or instruction but no external communication. None of the municipalities has organized their communication regarding the use of data science in a systematic way. Existing communication focuses on risks, especially on privacy and security.

The only municipality with experience stated they were forced to improve communication by circumstances: media, interest groups, and politics picked up on their use of ADM and started asking questions. The initial reactions from the municipality were ad hoc and defensive, as they were hesitant to give too much insight risking people "gaming the system". This only led to more attention and questions. Learning from this experience they realized that a systematic and proactive communication is needed. They also realized that some subjects are more sensitive than others. By initially focusing on detecting fraud and efficacy, instead of, for example, process efficiency, they received different attention and response. While they have shown that using ADM yielded significant improvements in efficiency and efficacy of fraud detection, the real value of fraud detection is inherently debatable (e.g., does effective fraud detection result in more or less fraud?).

Politicians have a significant role in gaining trust of the public by guarding the "human dimension" and preventing too technocratic communications. Accountability and communication are complementary. Being accountable is not "hiding behind the model", but explaining that a model was used to organize and analyze the information in the decision-making process.

All municipalities acknowledge that communication is about content, form and timing, and identifying target audiences. They tailor communication for three different target audiences for communication: laymen that seek general information, people more knowledgeable on the content, and technical experts. Municipalities feel that explaining and communicating the results of decision-making processes is (or should be) "business as usual" for public servants and politicians. But uncertainty and lack of expertise lead to an internal focus, neglecting external communication. This results in a lack of transparency instead of the professed openness. Municipalities signal the lack of a common language to communicate about using data science.

**COM2: Developments and plans for communication**

Ambitions regarding communication about the use of data science are still abstract and idealistic. Only one municipality reported that they had specific plans about designing and implementing communication systematically in the organization. Their ambition is to communicate as proactive as possible, using modern tools such as infographics and animations. They have also engaged external expertise to get help with this process.

The other municipalities have only thought about communication in more general terms. Their remarks can be clustered around three concepts:

On the purpose of communication:

- ADM can help achieve and illustrate the equal treatment of citizens.

- The more complex an ADM system, the less explainable it becomes.

- Municipalities should consider explicitly communicating both what they do as a part of their ADM processes as well as what they (intentionally) do not do. Communicating about intentions is important.

On the execution of communication:

- Form, content, and timing of the message should be tailored to the target audience.

- Who is communicating the message makes a difference in how the message is received.
- Communicating proactively often works better than communicating reactively.
- A common language and terminology is needed.

On the concept of trust:

- Trust is very important. Being transparent, engaging interest groups, and oversight by an independent third party are thought to help achieve trust.
- Politicians and other leaders play a key role in leading by example. They must have sufficient knowledge of ADM to be able to communicate effectively to facilitate trust.
- Explaining the process may create trust without having to explain the tools.

**PFM1: Current performance measurement**

Several municipalities have implemented some form of quality management, for example monitoring of citizen satisfaction. Data analysis may be used for policy analysis and substantiation.

The municipality using the algorithm for fraud detection has been able to quantify the effects, since the output variable (number of detected fraud cases) stayed the same. The major performance improvement however was not a change in the number of detected fraud cases, but a 95% reduction in the effort (in hours) needed to get similar results as before. Also 95% of the clients no longer had to submit forms, bank statements etc., so citizen satisfaction improved. This is one of the main goals of the administration. In this case the effect of using the algorithm could be determined and was soon evident. However, since the original intention of using the algorithm (as stated by the administration) was to increase the number of fraud cases detected, the performance did not live up to expectations since there was no substantial increase. In hindsight they realized this was more a problem of wrong communication. If they had stated their goal differently, negative attention may have been avoided. With investments in improved communication, they now aim to be transparent about the process that was followed, the inputs and the (expected and achieved) outputs.

For municipalities running pilot projects, their main focus is still justifying the investments in data science in terms of the achieved added value. This requires results to be quantified and monetized, preferably in an operational context. However, apart from the feasibility of such an operation, most data scientists lack the (business or financial) background for this.

**PFM2: Developments and plans for performance measurement**

Municipalities are, in general, not yet able to measure or evaluate the outcome of a policy or a process. They state that this requires investments that simply will not yield (enough) positive returns, either quantitative or qualitative. However, with more and more data sources becoming available, they expect data science to enable (better) measurement of outcomes in the future.

Municipalities also wonder about the norms and standards they should use when measuring ADM performance. Some are inherent in the science and the statistics, but municipalities are faced with the practical application of data science results. They explicitly state the need to know how to qualify the results as 'good' or 'bad'.

One municipality is looking into using a social cost-benefit analysis (CBA) to determine the impact and added value, of using ADM. They consider drafting a CBA with and without the algorithm, thus getting an insight in the differences.

**OTH1: Other aspects that could determine the impact of ADM on the quality of public services**

Finally, some general thoughts and remarks emerged from the interviews:

- In general, (citizen) trust is likely to improve because the ADM use increases across the society. In the future, the increase in trust to ADM may lead to a false sense of security.
- What competencies does the civil service of the future require? The municipalities need to raise awareness of their role in the information chain.
- At this stage, there are some important prerequisites for successful ADM implementation such as specific procurement processes, commissioning, and project management; both within the organization and in managing third parties.
- A national and overarching strategy for cooperation among municipalities seems evident.

V. DISCUSSION AND CONCLUSION

In this section, we discuss the results and provide actionable recommendations for various stakeholders of the local public administration processes in their journey of adopting ADM.

*On algorithms:* The adoption of ADM in governments is at early stages. There is a lack of ADM expertise in municipalities and a common terminology about algorithms and data science. In order to maximize the benefits and avoid the pitfalls, government organizations need to increase ADM maturity, not only at a technical level but also at managerial and operational level. The minimal maturity level to achieve initially should satisfy the necessary knowledge, skills, and infrastructure to define meaningful business cases for ADM, make informed decisions on how to use ADM responsibly, and manage outsourcing and procurement of systems that embody ADM components.

Making the distinction ADM types is important for the governments that are aiming at regulating the use of ADM. Algorithms that directly concern individuals or citizen groups require careful governance due to their potential impact.

*On governance of ADM:* There is consensus on the need for ADM regulation, preferably on a national level. However, at the early stage of ADM adoption, experimentation is required to learn. Thus, any regulation attempt should provide flexibility and a room for experimentation and innovation to develop maturity by learning.

Municipalities are careful and reserved in the use of ADM. They are primarily concerned about the risks and accountability. This leads to underutilization of ADM and lost opportunities to improve public services. Municipalities may benefit from a governance framework that provides guidance, tools, methods, and good practices. It is essential that the framework should not only address risks and problems, but also helps to identify the benefits and the business case of ADM.

Most municipalities work with third parties or procure off-the-shelf systems that embody ADM components. Such collaborations with third parties are not regulated in terms of ADM governance. In some cases, the municipality is not even aware that the third-party solutions utilize ADM, which creates additional complexities around the topic of accountability.

Umbrella organizations such as VNG can play a vital role in providing national guidelines for municipalities. By collecting information about cases of ADM use from municipalities, they can curate a set of good practices and a registry of potential risks.

*On communication:* Even though municipalities aspire to be transparent and open regarding the use of ADM, they do not have systematic plans to achieve this. They are forced to create ad hoc communication plans as a reaction to sudden negative political and media attention towards ADM. Instead, municipalities should create proactively communication strategies. However, this requires a certain level of maturity on ADM within municipalities and a shared language among all actors of government and citizens.

*On performance measurement:* The expectations from ADM and performance goals should be clearly defined as early as possible in the adoption phase. There is more to performance evaluation of ADM than merely measuring the accuracy, effectiveness, and efficiency. The impact of ADM on different groups and individuals should also be evaluated as well as on the quality of public service delivery. This depends on the testability and auditability of the ADM systems.

**Recommendations for practice**

<u>Umbrella organizations (e.g., VNG)</u>: Being representatives of municipalities, the umbrella organizations play a leading role in making sure that the municipal use of ADM benefits citizens. As an authority and advocate for municipalities these umbrella organizations can provide guidance to municipalities. The specific recommendations are as follows.

- Small(er) municipalities lack the expertise and resources to experiment with using ADM. Standalone experiments carry a high risk of failure for them. They could benefit from adopting tried and tested methods. The umbrella organization can cooperate with municipalities to provide these good practices.
- Sentiments in national politics and media regarding ADM are getting more tense which may lead to strict regulation. This can thwart the implementation of ADM by municipalities. The umbrella organization should facilitate an optimal environment for municipalities to use data science. That means the opportunity to test and experiment.
- Most municipalities expect external guidance, in addition to laws and regulations. The umbrella organization can provide templates for governance and communication.

<u>Individual municipalities</u>: Developing a sufficient level of expertise within the municipal organization is imperative to cope with the challenges of ADM adoption. The specific recommendations are as follows.

- Raise ADM literacy and awareness of risks and benefits of ADM at the management level.
- Actively involve senior management, policy makers, and business controllers in ADM processes to better assess and define the business value of using ADM.
- Only engage with third parties under proper regulation. Certain level of ADM knowledge is required to deal with third parties and manage the collaboration.
- To gain trust, focus on explaining your intentions.
- Exercise proactive communication. Do not wait until a crisis to take place.

**Limitations and future work:**

The maturity and scope of ADM use in municipalities is still limited. Therefore, our findings only capture the early stages of ADM adoption by municipalities. When the use of algorithms become more widespread, more in-depth research can be done. It would be ideal to repeat the case study after several years, to observe how the use of ADM evolves.

Even though we selected the interviewed municipalities to be as representative of the population as possible, this study is still limited due to the number of cases; 12 interviews from eight municipal organizations.

Our results clearly indicate the need for a governance framework for ADM use in public sector. Building on the results of this study, we recommend the development of a framework and evaluation of it in real-life use.

ACKNOWLEDGMENT

We would like to express our gratitude VNG for their support in this study.

REFERENCES

[1] S. Newell and M. Marabelli, 'Strategic opportunities (and challenges) of algorithmic decision-making: A call for action on the long-term societal effects of "datification"', *The Journal of Strategic Information Systems*, vol. 24, no. 1, pp. 3–14, 2015.

[2] S. Garfinkel, J. Matthews, S. S. Shapiro, and J. M. Smith, *Toward algorithmic transparency and accountability*. ACM New York, NY, USA, 2017.

[3] J. Baijens, T. Huygh, and R. Helms, 'Establishing and theorising data analytics governance: a descriptive framework and a VSM-based view', *Journal of Business Analytics*, pp. 1–22, Jul. 2021, doi: 10.1080/2573234X.2021.1955021.

[4] V. Grover, R. H. L. Chiang, T.-P. Liang, and D. Zhang, 'Creating Strategic Business Value from Big Data Analytics: A Research Framework', *Journal of Management Information Systems*, vol. 35, no. 2, pp.

388–423, Apr. 2018, doi: 10.1080/07421222.2018.1451951.
[5] B. Lepri, N. Oliver, E. Letouzé, A. Pentland, and P. Vinck, 'Fair, transparent, and accountable algorithmic decision-making processes', *Philosophy & Technology*, vol. 31, no. 4, pp. 611–627, 2018.
[6] B. J. Dietvorst, J. P. Simmons, and C. Massey, 'Algorithm aversion: People erroneously avoid algorithms after seeing them err.', *Journal of Experimental Psychology: General*, vol. 144, no. 1, p. 114, 2015.
[7] KPMG, 'Vertrouwen van de Nederlandse burger in Algoritmes'. 2019. [Online]. Available: https://kennisopenbaarbestuur.nl/media/257150/vertrouwen-van-de-nederlandse-burger-in-algoritmes.pdf
[8] M. Kuziemski and G. Misuraca, 'AI governance in the public sector: Three tales from the frontiers of automated decision-making in democratic settings', *Telecommunications policy*, vol. 44, no. 6, p. 101976, 2020.
[9] L. Andrews, 'Public administration, public leadership and the construction of public value in the age of the algorithm and "big data"', *Public Administration*, vol. 97, no. 2, pp. 296–310, 2019.
[10] P. K. Agarwal, 'Public administration challenges in the world of AI and Bots', *Public Administration Review*, vol. 78, no. 6, pp. 917–921, 2018.
[11] G. Barber, 'Artificial Intelligence Confronts a 'Reproducibility' Crisis', *Wired, September*, 2019, [Online]. Available: https://www.wired.com/story/artificial-intelligence-confronts-reproducibility-crisis/
[12] M. Veale, M. Van Kleek, and R. Binns, 'Fairness and accountability design needs for algorithmic support in high-stakes public sector decision-making', in *Proceedings of the 2018 chi conference on human factors in computing systems*, 2018, pp. 1–14.
[13] R. Brauneis and E. P. Goodman, 'Algorithmic transparency for the smart city', *Yale JL & Tech.*, vol. 20, p. 103, 2018.
[14] A. Brown, A. Chouldechova, E. Putnam-Hornstein, A. Tobin, and R. Vaithianathan, 'Toward algorithmic accountability in public services: A qualitative study of affected community perspectives on algorithmic decision-making in child welfare services', in *Proceedings of the 2019 CHI Conference on Human Factors in Computing Systems*, 2019, pp. 1–12.
[15] C. Tobler, *Limits and potential of the concept of indirect discrimination*. Office for Official Publications of the European Communities, 2008.
[16] B. Aysolmaz, D. Iren, and N. Dau, 'Preventing Algorithmic Bias in the Development of Algorithmic Decision-Making Systems: A Delphi Study', 2020.
[17] L. Edwards and M. Veale, 'Slave to the algorithm: Why a right to an explanation is probably not the remedy you are looking for', *Duke L. & Tech. Rev.*, vol. 16, p. 18, 2017.
[18] F. Hendriks, 'Understanding good urban governance: Essentials, shifts, and values', *Urban Affairs Review*, vol. 50, no. 4, pp. 553–576, 2014.
[19] A. Meijer, M. T. Schäfer, and M. Branderhorst, 'Principes voor goed lokaal bestuur in de digitale samenleving: Een aanzet tot een normatief kader', *Bestuurswetenschappen*, vol. 73, no. 4, pp. 8–23, 2019.
[20] N. Diakopoulos, 'Accountability in algorithmic decision making', *Communications of the ACM*, vol. 59, no. 2, pp. 56–62, 2016.
[21] N. Diakopoulos, 'Accountability in Algorithmic Decision-making: A view from computational journalism', *Queue*, vol. 13, no. 9, pp. 126–149, 2015.
[22] H. Shah, 'Algorithmic accountability', *Philosophical Transactions of the Royal Society A: Mathematical, Physical and Engineering Sciences*, vol. 376, no. 2128, p. 20170362, 2018.
[23] D. Doneda and V. A. Almeida, 'What is algorithm governance?', *IEEE Internet Computing*, vol. 20, no. 4, pp. 60–63, 2016.
[24] B. Waltl and R. Vogl, 'Increasing Transparency in Algorithmic-Decision-Making with Explainable AI', *Datenschutz und Datensicherheit-DuD*, vol. 42, no. 10, pp. 613–617, 2018.
[25] B. Goodman and S. Flaxman, 'European Union regulations on algorithmic decision-making and a "right to explanation"', *AI magazine*, vol. 38, no. 3, pp. 50–57, 2017.
[26] S. Wachter, B. Mittelstadt, and L. Floridi, 'Why a right to explanation of automated decision-making does not exist in the general data protection regulation', *International Data Privacy Law*, vol. 7, no. 2, pp. 76–99, 2017.
[27] S. Doove and D. Otten, *Verkennend onderzoek naar het gebruik van algoritmen binnen overheidsorganisaties*. Retrieved from Den Haag: https://www. cbs. nl/nlnl/maatwerk/2018/48/gebruik …, 2018.
[28] M. Ausloos, R. Cerqueti, and T. A. Mir, 'Data science for assessing possible tax income manipulation: The case of Italy', *Chaos, Solitons & Fractals*, vol. 104, pp. 238–256, Nov. 2017, doi: 10.1016/j.chaos.2017.08.012.
[29] R. Fujdiak, P. Masek, P. Mlynek, J. Misurec, and E. Olshannikova, 'Using genetic algorithm for advanced municipal waste collection in Smart City', in *2016 10th International Symposium on Communication Systems, Networks and Digital Signal Processing (CSNDSP)*, Prague, Czech Republic, Jul. 2016, pp. 1–6. doi: 10.1109/CSNDSP.2016.7574016.
[30] K. Levy, K. E. Chasalow, and S. Riley, 'Algorithms and Decision-Making in the Public Sector', *Annu. Rev. Law. Soc. Sci.*, vol. 17, no. 1, p. annurev-lawsocsci-041221-023808, Oct. 2021, doi: 10.1146/annurev-lawsocsci-041221-023808.
[31] G. Andrienko *et al.*, '(So) Big Data and the transformation of the city', *Int J Data Sci Anal*, vol. 11, no. 4, pp. 311–340, May 2021, doi: 10.1007/s41060-020-00207-3.
[32] R. K. Yin, 'Case study research: Design and methods, applied social research', *Methods series*, vol. 5, 1994.
[33] M. Saunders, P. Lewis, and A. Thornhill, 'Research methods', *Business Students 4th edition Pearson Education Limited, England*, 2007.

# APPENDIX - I

TABLE III.    INTERVIEW QUESTIONS

| CODE | QUESTIONS |
|---|---|
| Background | - What is your function?<br>- What are your responsibilities, and who do you report to?<br>- (How) are you involved with ADM / data science?<br>- Does your municipality have a data lab or data science department? How is it organized? |
| ALG1 | - In general, is the municipality actively using algorithms / data science?<br>- Do you have a defined process for data science/ADMs?<br>- Are there any examples? |
| ALG1.1 | - Who (or which departments) are involved? Please elaborate (operational, responsibilities, third parties, etc.)<br>- Who (or which departments) are responsible for data management?<br>- In which policy fields is ADM exercised? |
| ALG1.2 | - What are the considerations for using ADM?<br>- What drives the decision to use or not use ADM?<br>- Do you know of instances where existing ADM was explicitly put out of use, or the decision to use new ADM was explicitly rejected? |
| ALG1.3 & 1.4 | - What benefits and risks of using ADM were identified in your organization?<br>- Has there been a systematic risk/benefit analysis? |
| ALG2 | - What developments regarding the use of ADMs do you see in the (near) future? |
| GOV1 | - What is the current scope of IT/data governance?<br>- Does the current IT governance framework cover ADM practices in particular? |
| GOV1.1 | - Are there collaborations with third parties (e.g., software providers, consultants) which lead to the (explicit or implicit) use of ADM?<br>- If so, how are such arrangements and contracts managed and controlled? |
| GOV1.2 | - How is the senior management and the municipal council involved in the governance of ADM?<br>- What is their role in the acquisition, implementation, execution, and communication of ADM? |
| GOV2 | - What are the plans for establishing governance mechanisms for ADM?<br>- What is the involvement of higher government organizations in the governance of ADM at municipal levels? |
| POL1 | - What are the specific policy fields in which there are instances of ADM use?<br>- Who is involved in these policies and how?<br>- (How) are the policy fields impacted by the use of ADM? |
| COM1 | - How and when the communication occurs about the use of the algorithm, both internally and externally?<br>- Has the employees received any training on ADM or data science?<br>- (How) are citizens informed about the use of algorithms?<br>- What form of communication is used (e.g., disclaimers, website info, letters) |
| COM2 | - If any, what are the plans for establishing a communication policy regarding the use of ADM? |
| PFM1 | - How is the performance of the process measured?<br>- How were the goals of ADM set? |
| PFM2 | - What are the plans for setting a systematic performance measurement procedure in place? |